\documentclass[prx,twocolumn,amsmath,amssymb,superscriptaddress,floatfix,nofootinbib,aps]{revtex4-2}

\usepackage{amsmath}
\usepackage{amsfonts}
\usepackage{graphicx}
\usepackage{dcolumn}
\usepackage{bm}
\usepackage{amsthm}
\usepackage{algorithm}
\usepackage{algpseudocode}
\usepackage[cmyk]{xcolor}
\usepackage[colorlinks,bookmarks=false,citecolor=magenta,linkcolor=magenta,urlcolor=magenta]{hyperref}
\usepackage{braket}
\usepackage{orcidlink}

\begin{document}

\preprint{APS/123-QED}
\title{Phase diagram of a dual-species Rydberg atom ladder}

\author{Lei-Yi-Nan Liu \orcidlink{0009-0005-3072-3650}}
 \affiliation{%
 School of Physics, Beihang University, Beijing 100191, China
}

\author{Shi-Rong Peng}
 \affiliation{%
 School of Physics, Beihang University, Beijing 100191, China
}

\author{Ze-Yuan Huang}%
 \affiliation{%
 School of Physics, Beihang University, Beijing 100191, China
}

\author{Xing-Man Wei}%
 \affiliation{%
 School of Physics, Beihang University, Beijing 100191, China
}

\author{Yun-Han Zou}%
 \affiliation{%
 School of Physics, Beihang University, Beijing 100191, China
}

\author{Su Yi}
\affiliation{Institute of Fundamental Physics and Quantum Technology \& School of Physical Science and Technology, Ningbo University, Ningbo, 315211, China}
\affiliation{Peng Huanwu Collaborative Center for Research and Education, Beihang University, Beijing 100191, China}

\author{Jian Cui \orcidlink{0000-0001-6643-7625}}
 \email{jiancui@buaa.edu.cn}
 \affiliation{%
 School of Physics, Beihang University, Beijing 100191, China
}
\date{\today}

\begin{abstract}
Dual-species Rydberg atom arrays extend single-species platforms by introducing competing interaction scales and enhanced quantum fluctuations, enabling phenomena beyond homogeneous settings. In this work, we study the ground-state phase diagram of a one-dimensional dual-species Rydberg atom ladder using large-scale density-matrix renormalization group calculations. We identify disordered phases, multiple ordered phases with $\mathbb{Z}_2$, $\mathbb{Z}_3$, and $\mathbb{Z}_4$ symmetry, as well as floating phases characterized by incommensurate wave vectors and algebraically decaying correlations. 
Importantly, we observe a smooth crossover between distinct $\mathbb{Z}_2$-ordered regimes, reflecting a reorganization of low-energy degrees of freedom rather than a true phase transition, which is absent in single-species Rydberg arrays. We further uncover a multi-critical point at the boundary between the $\mathbb{Z}_2 \otimes \mathbb{Z}_2$ and $\mathbb{Z}_3 \otimes \mathbb{Z}_3$ ordered phases, where Ising, chiral, and first-order transition lines intersect. Our results demonstrate that dual-species Rydberg atom arrays provide a unique platform for realizing crossover physics and multi-critical behavior inaccessible in single-species architectures.
\end{abstract}

\maketitle
\section{introduction}
Rydberg atom arrays are powerful and highly versatile platforms 
for investigating strongly correlated quantum many-body physics. 
Owing to their intrinsically long-range interactions, high degree of tunability, and 
direct experimental accessibility, these systems provide an unprecedented level of control over 
both microscopic Hamiltonian parameters and lattice geometry \cite{Henriet2020,Browaeys2020}. 
By trapping and manipulating individual neutral atoms in programmable optical tweezer arrays, 
a wide variety of effective lattice models can be realized with single-site resolution and 
flexible dimensionality \cite{Schneider2012,Martinez2017}. 
A central feature of Rydberg systems is the strong van der Waals 
interaction between excited atoms, which gives rise to the Rydberg blockade effect \cite{Urban2009,Gaetan2009}. 
This blockade enforces local constraints on the accessible Hilbert space, 
effectively suppressing simultaneous 
excitations within a characteristic blockade radius and leading to constrained many-body dynamics. 
As a result, Rydberg atom arrays naturally host a wealth of collective phenomena, including 
commensurate density-wave ordering \cite{Schaub2012,Schaub2015,Labuhn2016,Bernien2017}, 
versatile quantum phase transitions \cite{Bernien2017,O'Rourke2023,Samajdar2020,Rhine2021,Zhang2025}, 
non-thermal dynamics \cite{Bernien2017,Turner2018,HoWenWei2019,Choi2019}, 
and topologically ordered quantum states \cite{Verresen2021,Satzinger2021}, 
making them an ideal setting for exploring quantum many-body physics beyond 
conventional solid-state platforms.

In single-species Rydberg atom arrays, the resulting many-body physics is by now relatively well understood.
In one dimension, the ground-state phase diagram has been extensively investigated and is known to 
host a sequence of ordered phases with different patterns of translational symmetry breaking, as well as floating phases characterized by incommensurate wave vectors and algebraically decaying correlations \cite{Bernien2017,Michael2019,Zhang2025}. 
The nature of the phase transitions between these phases has also been extensively investigated.
Transitions from disordered to ordered phases can belong to the Ising, Potts, Ashkin-Teller (AT), or chiral universality classes, or proceed via a Kosterlitz-Thouless (KT) transition followed by a Pokrovsky-Talapov (PT) transition through an intermediate floating phase \cite{Yu2022,Maceira2022,Michael2019}.
Beyond one dimension, single-species Rydberg systems have also been studied in a variety of 
lattice geometries, such as ladder configurations \cite{Soto-Garcia2025,Liao2025,Tsitsishvili2022,Fromholz2022}, 
square lattice \cite{Samajdar2020,Han2025}, triangular lattice \cite{Chang-Xiao2022}, 
honeycomb lattice \cite{Yang2022,Kornjaca2023}, Kagome lattice \cite{Rhine2021}, and Lieb lattice \cite{Mark2025},
revealing rich but conceptually well-established phase diagrams controlled by geometric 
frustration and blockade-induced constraints. 
In these homogeneous settings, the dominant physical phenomena can typically be understood in terms of a single interaction scale and a single blockade radius.

However, with ongoing advances in experimental techniques, Rydberg atom arrays have recently been 
extended beyond single-species implementations to realize dual-species architectures, 
in which two distinct atomic species or internal states coexist within 
the same lattice \cite{Anand2024,Zeng2017,Sheng2022,Singh2022}. 
In such systems, atoms belonging to different species can exhibit different interaction 
strengths and blockade radii, leading to a hierarchy of competing length and energy scales. 
This additional degree of freedom substantially enriches the structure of the effective many-body 
Hamiltonian and enables forms of frustration and competition that are absent in homogeneous arrays \cite{Lei2024}. 
As a consequence, the resulting collective behavior can no longer be understood solely in terms of a single 
blockade constraint, as is often sufficient for describing the dominant physics of single-species Rydberg systems. 
Instead, dual-species Rydberg arrays provide a fertile setting for realizing qualitatively 
new many-body phenomena, such as novel order-disorder mixture phase, tri-criticality and super-symmetry  \cite{Lei2024,Li2024}, 
thereby opening new avenues for exploring strongly 
correlated quantum matter in highly controllable atomic platforms.
Beyond their relevance for quantum simulation, dual-species Rydberg arrays also open new possibilities for quantum information processing. In particular, the presence of multiple interaction channels allows for more flexible encoding of quantum degrees of freedom and enables the implementation of generalized constraint models beyond the standard blockade regime \cite{Anand2024,Lei2024}. Importantly, the two species can be individually addressed and controlled, providing an additional layer of programmability compared to single-species platforms. This species-resolved control enables selective manipulation of interactions and local degrees of freedom, thereby allowing for the implementation of more complex quantum operations and circuit architectures \cite{Ryan2026,Francesco2026,Bikun2025}. As a result, dual-species systems offer enhanced flexibility for encoding qubits and designing quantum protocols. 

Despite these promising developments, the phase structure of dual-species 
Rydberg systems remains far less explored, particularly beyond strictly one-dimensional geometries. 
Motivated by these considerations, we previously explored the phase diagram of one-dimensional dual-species Rydberg atom arrays, revealing a range of ordered and disordered phases driven by competing interactions \cite{Lei2024}. Based on this, we extend the study to a ladder geometry, which represents the minimal setting beyond strictly one dimension. The ladder configuration introduces additional spatial structure and inter-chain couplings, further enhancing the interplay between competing interactions. As we demonstrate below, this seemingly modest extension leads to a qualitatively richer phase diagram, featuring crossover behavior and a multi-critical point that are absent in purely one-dimensional single-species or dual-species chains. The ladder geometry therefore provides a natural and experimentally relevant platform for uncovering new collective phenomena unique to multi-species Rydberg systems.

\section{Model}
\begin{figure}
    \centering
    \includegraphics[width=\linewidth]{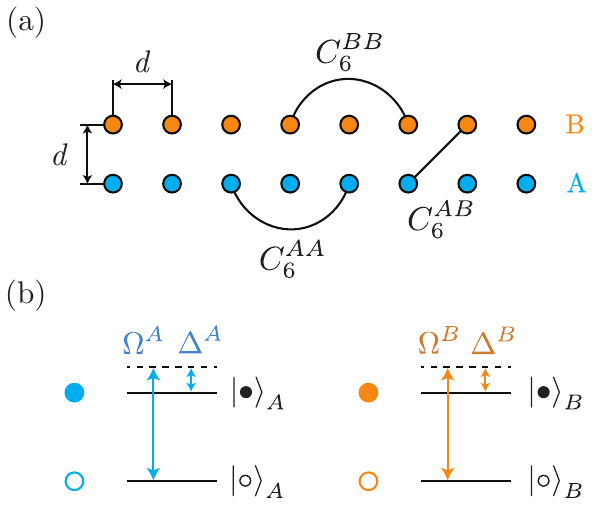}
    \caption{(a) Ladder lattice of two-species Rydberg atoms. Blue (A) and orange (B) circles denote the two atomic species, with lattice spacing $d$ along both legs and rungs. The van der Waals interactions are characterized by $C_6^{AA}$, $C_6^{AB}$, and $C_6^{BB}$. (b) Illustration of energy levels, detunings $\Delta$ and external fields $\Omega$ of A and B atoms.
    }
    \label{model}
\end{figure}

We consider a two-species Rydberg atom system arranged in a ladder geometry, as illustrated in Fig.~\ref{model}(a). Each lattice site is occupied by either a type-A or type-B atom, denoted by blue and orange circles, respectively, with lattice spacing $d$ along both the legs and rungs. Both species are coherently driven to their Rydberg states with Rabi frequencies $\Omega^{A,B}$ and detunings $\Delta^{A,B}$. For simplicity, throughout this work we assume identical Rabi frequencies and detunings for the two species, i.e., $\Omega^A=\Omega^B\equiv\Omega$ and $\Delta^A=\Delta^B\equiv\Delta$. The atoms interact via van der Waals interactions characterized by the $C_6$ coefficients $C_6^{AA}$, $C_6^{BB}$, and $C_6^{AB}(=C_6^{BA})$. As shown in Fig.\ref{model}(c), we use $\ket{\circ}$ and $\ket{\bullet}$ to denote the ground and Rydberg states for both atomic species.

The Hamiltonian of the system is given by 
\begin{eqnarray}
    \hat{H} &=& \hat{H}_0+\hat{H}_1, \label{H}\\
    \hat{H}_0 &=& \sum_{s=A,B}\sum_{i\in s}\left(\frac{\Omega^s}{2}\hat{\sigma}_i^x-\Delta^s\hat{n}_i\right), \label{H0}\\
    \hat{H}_1 &=& \sum_{\substack{s=A,B\\ s'=A,B}}\sum_{\substack{i<j\\i\in s,j\in s'}}\frac{C_6^{ss'}}{|\boldsymbol{r}_{ij}|^6}\hat{n}_i\hat{n}_j, \label{H1}
\end{eqnarray}
where $H_0$ and $H_1$ denote the single-site driving term and the interaction term, respectively.
The operator $\hat{\sigma}^x_i = \ket{\circ}_i\bra{\bullet}+\ket{\bullet}_i\bra{\circ}$ is the Pauli operator in $x$-direction and $\hat{n}_i=\ket{\bullet}_i\bra{\bullet}=(\mathbb{I}-\hat{\sigma}_i^z)/2$ is the Rydberg excitation number operator on site $i$. The distance between atoms $i$ and $j$ is denoted as $|\boldsymbol{r}_{ij}|=|\boldsymbol{r}_i-\boldsymbol{r}_j|$ with $\boldsymbol{r}_i(\boldsymbol{r}_j)$ the position of atom $i$($j$). 
Following the convention in single-species Rydberg systems, we introduce an effective length scale $R_b = \sqrt[6]{|C_6^{AB}|/\Omega}$ to characterize the competition between the external driving $\Omega$ and the interatomic interactions. 
Here, we use $(C_{AA}, C_{BB}, C_{AB})=(220.79, 38.84, -4.2)\ \mathrm{GHz}\cdot\mu \mathrm{m}^6$ 
throughout this work, which corresponds to the interactions between $\ket{64S_{1/2}}$ of Caesium (A) atom and 
$\ket{54S_{1/2}}$ of Rubidium (B) atom \cite{Sibalic2017}. 


\begin{figure*}
    \includegraphics[width=1\linewidth]{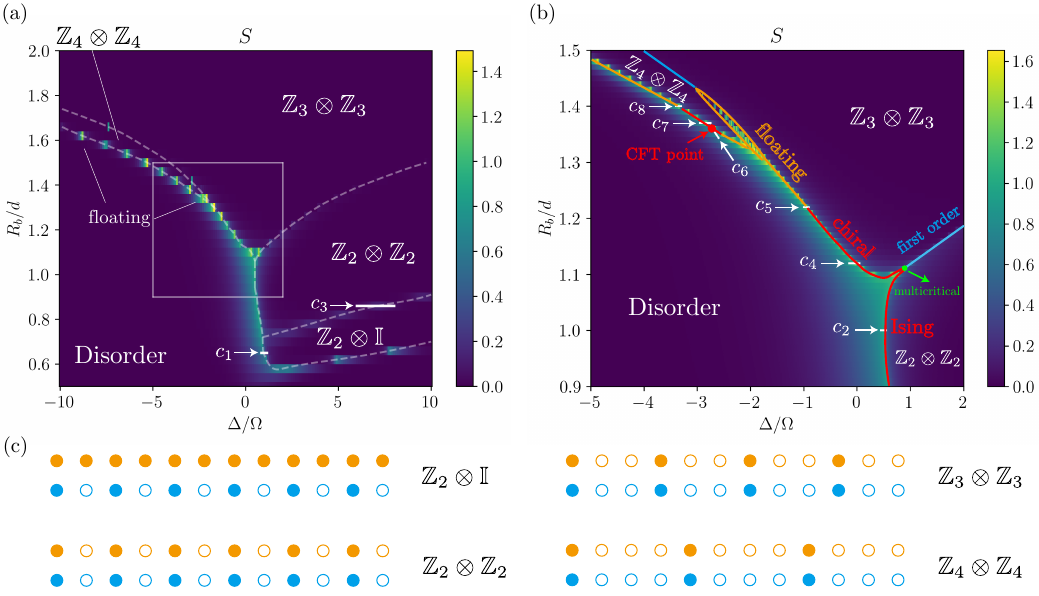}
    \caption{
    (a) Ground-state phase diagram of the ladder system. Four distinct ordered phases are identified, as illustrated in (c). The phase boundaries are determined from the bipartite entanglement entropy $S$ and are indicated by white dashed lines. 
    (b) Enlarged view of the region highlighted by the white square in (a). This regime contains three ordered phases, one disordered phase, and several floating-phase regions. Blue (red) lines denote first-order (continuous) phase transitions. Floating phases are marked by or enclosed within orange lines. The green dot indicates the multi-critical point, while the red dot marks a CFT point belonging to the Ashkin-Teller universality class. Among the continuous transition lines (red), all except the one labeled Ising correspond to chiral transitions. Representative parameter cuts analyzed in the following sections are also indicated.
    (c) Schematic configurations of the ordered phases shown in (a) and (b).
    }
    \label{phase-diagram}
\end{figure*}


\section{Methods}
\subsection{Numerical method}

We compute the phase diagram of the aforementioned model using the density matrix renormalization group 
(DMRG) method \cite{SRWhite1992,SRWhite1993,Ulrich2011} implemented in ITensor \cite{ITensor,ITensor-r0.3}. 
Unless otherwise specified, open boundary conditions are employed. In our calculations, interaction terms with energy scales larger than $10^{-10}\lvert C_6^{AB}\rvert / d^6$ are retained, corresponding to an effective interaction cutoff length of $10^{10/6} \approx 46$, which ensures sufficient numerical accuracy. The maximum bond dimension is set to $512$, and the truncation error is controlled to be below $10^{-11}$. The length of the ladder is denoted by $L$, and the total number of atoms is therefore $N = 2L$.

\subsection{Entanglement entropy}
In this paper, we primarily use the bipartite entanglement entropy $S$ as a probe of quantum phase transitions. Unlike conventional diagnostics such as order parameters or energy gaps, entanglement entropy provides a more unbiased characterization of criticality: the identification of suitable order parameters generally requires a priori knowledge of the underlying phase ordering, while computing energy gaps can be computationally demanding in large systems \cite{Samajdar2020}. 
We define the bipartite entanglement entropy of the ground state as
\begin{equation}
    S(\rho_{\ell}) = -\mathrm{tr}\left(\rho_{\ell}\log_2 \rho_{\ell}\right),
\end{equation}
where $\rho_{\ell}$ is the reduced density matrix of a subsystem consisting of the first $\ell$ rungs, obtained by cutting the ladder along a bond parallel to the rung direction. 

If the critical point is described by conformal field theory (CFT), the bipartite entanglement entropy is predicted to scale as \cite{Calabrese2009,Calabrese2004}
\begin{equation}
    S(\rho_{\ell}) \sim \frac{c}{6}\log_2\left(
    \frac{2L}{\pi}\sin\frac{\pi\ell}{L}
    \right), 
\end{equation}
from which the central charge $c$ can be extracted by fitting. 
The value of the central charge allows us to distinguish different critical points and identify floating phases.

\subsection{Order parameter and Binder cumulant}
We first define the average density operator on the $i$-th rung as
\begin{equation}
    \hat{n}_i^r=\frac{\hat{n}_i^A+\hat{n}_i^B}{2}
\end{equation}
where $\hat{n}_i^{A(B)}$ denotes the excitation-number operator of the A(B) atom on rung $i$. 
The rung density is then the average excitation density of atom A and B. 

Based on the rung density operator $\hat{n}^r_i$, we can define the order parameter for a $p$-periodic ordered phase as 
\begin{equation}
    \hat{O}_p(t, L) = \frac{1}{L}\sum_{j=1}^L e^{i\frac{2\pi}{p}j}\hat{n}_j^r, 
\end{equation}
where $t$ is a tuning parameter in the Hamiltonian and the corresponding Binder cumulant \cite{Binder1981,Binder198106} is, 
\begin{equation}
    U_p(t, L) = 1-\frac{\langle|\hat{O}_p|^4\rangle}{3\langle|\hat{O}_p|^2\rangle^2},
\end{equation}
where $\langle|\hat{O}_p|^2\rangle\equiv\langle\hat{O}_p^{\dagger}\hat{O}_p\rangle$ and $\langle|\hat{O}_p|^4\rangle\equiv\langle\hat{O}_p^{\dagger}\hat{O}_p\hat{O}_p^{\dagger}\hat{O}_p\rangle$. 
The Binder cumulant is a dimensionless quantity that probes the fluctuations of the order parameter and provides a sensitive indicator of criticality. For continuous phase transitions, curves of $U_p(t,L)$ for different system sizes intersect at the critical point, allowing for an accurate determination of the phase boundary. 

These definitions can be analogously applied to the A and B sub-lattices. For example, 
\begin{eqnarray}
    \hat{O}_p^{A(B)} &=& \frac{1}{L}\sum_{j=1}^L e^{i\frac{2\pi}{p}j}\hat{n}_j^{A(B)}, \\
     U_p^{A(B)} &=& 1 - \frac{\langle|\hat{O}_p^{A(B)}|^4\rangle}{3\langle|\hat{O}_p^{A(B)}|^2\rangle^2}. 
\end{eqnarray}
The order parameter and Binder cumulant further allow us to extract the critical exponents $\nu, \beta$ and the position of critical point $t_c$ via the finite-size scaling relations
\begin{eqnarray}
    |\langle\hat{O}_p(t, L)\rangle| &=& L^{-\beta/\nu} f_1\left[L^{1/\nu}(t-t_c)\right], \\
    U_p(t, L) &=& f_2\left[L^{1/\nu}(t-t_c)\right], 
\end{eqnarray}
where $f_1$ and $f_2$ are universal scaling functions independent of system size. 

\subsection{Correlation function and phase transition indicator}
To extract the correlation length $\xi$ and the wave vector $q$ (defined in units of $2\pi$) from finite-size DMRG calculations, we analyze the behavior of correlation functions. We define the density-density correlation function as
\begin{equation}
    C_s(x)=C(s, x) := \langle \hat{n}_s^r \hat{n}_{s+x}^r\rangle - \langle \hat{n}_s^r\rangle\langle \hat{n}_{s+x}^r\rangle, 
\end{equation}
where $s$ denotes a fixed reference site. The corresponding structure factor is then defined as 
\begin{equation}
    S_s(k) = \sum_{x=1}^{L} e^{ik x} C_s(x), 
\end{equation}
with $k = 2\pi q$. In practice, we choose $s$ to be the central site of the system to minimize boundary effects. Exploiting the inversion symmetry of the system, the structure factor simplifies to 
\begin{equation}
    S_s(k) = \sum_{x=1}^{L} C_s(x) \cos(k x). 
\end{equation}
For completeness, we also introduce the structure factor constructed from all 
pairwise density correlations in the system, 
\begin{equation}
    S(k) = \frac{1}{L}\sum_{x, y=1}^Le^{ik(x-y)}C(x,y), 
\end{equation}
which provides a global characterization of density fluctuations in momentum space.

In one-dimensional systems, the correlation function generally follows the Ornstein--Zernike (OZ) form
\begin{equation}
    C(x)\sim A_0\frac{e^{-x/\xi}}{\sqrt{x}}\cos\left(2\pi qx+\phi_0\right).
\end{equation}
Although our system is defined on a ladder geometry, it remains quasi-one-dimensional due to its finite width. As a result, the long-distance behavior of correlation functions is still governed by an effective one-dimensional description, and the OZ form remains applicable. By fitting the finite-size DMRG data to this form, we extract both the correlation length $\xi$ and the wave vector $q$.

To distinguish different scenarios of commensurate-incommensurate (C-IC) phase transitions in the phase diagram, we introduce the dimensionless quantity $\xi \lvert q - 1/p \rvert$ as a diagnostic indicator. Near a C-IC transition, the indicator exhibits qualitatively distinct behaviors: it may diverge, approach a finite nonzero constant, or vanish, corresponding respectively to a Kosterlitz-Thouless (KT) transition into a floating phase, a chiral transition, and a conformal transition \cite{Huse1982,Maceira2022}.

\section{Phase Diagram}
\subsection{Overview}
Using DMRG simulations on a ladder of length $L=121$ ($N=2L=242$ sites), we obtain the phase diagram 
in the $\Delta/\Omega-R_b/d$ plane, as shown in Fig.~\ref{phase-diagram} (a). 
To distinguish different phases, we use the bipartite entanglement entropy $S=-\mathrm{Tr}(\rho_r \ln \rho_r)$ to map out the phase diagram, 
where $\rho_r$ is the reduced density matrix of half of the system.


In this phase diagram, we find four kinds of ordered phase, labeled by 
$\mathbb{Z}_2\otimes\mathbb{I}$, $\mathbb{Z}_2\otimes \mathbb{Z}_2$, $\mathbb{Z}_3\otimes \mathbb{Z}_3$ 
and $\mathbb{Z}_4\otimes \mathbb{Z}_4$, and their configuration is illustrated in Fig.~\ref{phase-diagram} (a). 
We use $\mathbb{Z}_n\otimes \mathbb{Z}_m$ to denote a periodicity of $n$ between type-A atoms, 
and $m$ between type-B atoms, respectively. 
The phase transitions between disordered phase and ordered phases are of second order while 
the transitions between $\mathbb{Z}_4\otimes \mathbb{Z}_4$ and $\mathbb{Z}_3\otimes \mathbb{Z}_3$, 
$\mathbb{Z}_3\otimes \mathbb{Z}_3$ and $\mathbb{Z}_2\otimes \mathbb{Z}_2$ are of first order. 
More interestingly, the boundary between $\mathbb{Z}_2\otimes \mathbb{Z}_2$ and $\mathbb{Z}_2\otimes \mathbb{I}$ is a crossover instead of a phase transition, as they show identical $\mathbb{Z}_2$ symmetry 
just like the $U(1)$ symmetry in the BEC-BCS crossover. 
There are also floating phase areas in the phase diagram, but it is too small to 
show explicitly here. 

In Fig.~\ref{phase-diagram} (b), we summarize the essential features of the phase diagram, 
excluding the crossover between the two distinct $\mathbb{Z}_2$-ordered phases. 
The floating phases are clearly identified, appearing at the boundaries of the 
$\mathbb{Z}_3$ and $\mathbb{Z}_4$ ordered regions. Notably, unlike in the single-species case, 
we do not observe a conformal critical point of the three-state Potts type separating the disordered phase 
from the $\mathbb{Z}_3 \otimes \mathbb{Z}_3$ ordered phase. Instead, the chiral transition line between 
the disordered and $\mathbb{Z}_3 \otimes \mathbb{Z}_3$ phases intersects with the Ising transition line, 
forming a multi-critical point in the phase diagram. 
Such a structure does not arise in single-species Rydberg atom arrays. 

\subsection{Ising transitions}
\begin{figure}
    \includegraphics[width=1\linewidth]{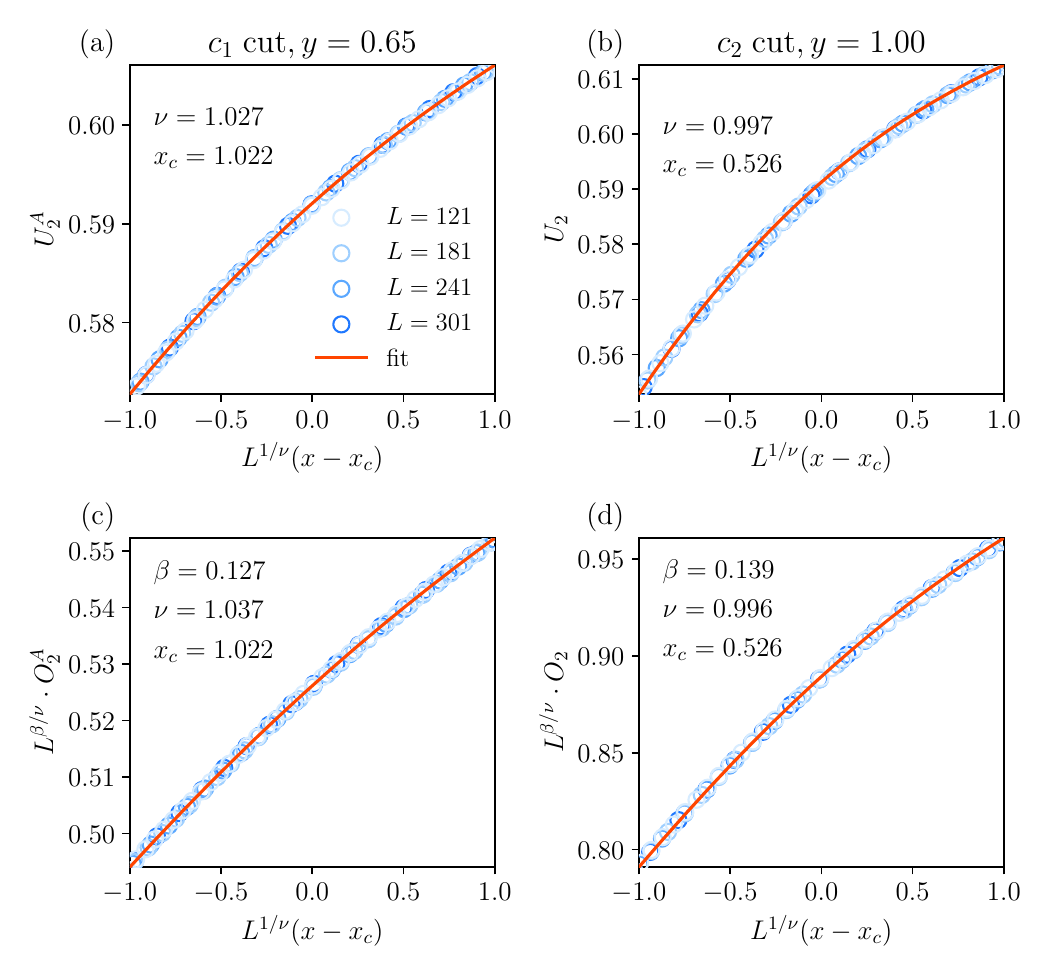}
    \caption{
    Finite size scaling analysis of binder cumulant and order parameter at cut $c_1$ and $c_2$. 
    We extract critical exponents $\beta, \nu$ and the position of critical point $x_c$ for $c_1$ cut 
    in (a)(c), and for $c_2$ cut in (b)(d). We fit the data using a 14th-order polynomial 
    by minimizing the distance between the data points and the polynomial curve. 
    }
    \label{ising_transition}
\end{figure}

In the regime of small $R_b/d$, increasing $\Delta/\Omega$ drives the system from a 
commensurate disordered phase into $\mathbb{Z}_2$-ordered phases. Through finite-size scaling analysis, 
we establish that the transitions from the disordered phase to both the $\mathbb{Z}_2 \otimes \mathbb{I}$ 
and $\mathbb{Z}_2 \otimes \mathbb{Z}_2$ phases belong to the Ising universality class. As shown in Fig.~\ref{ising_transition}, the $c_1$ ($c_2$) cut crosses the phase boundary between the disordered phase 
and the $\mathbb{Z}_2 \otimes \mathbb{I}$ ($\mathbb{Z}_2 \otimes \mathbb{Z}_2$) phase. 
The extracted critical exponents, $\nu$ and $\beta$, are close to the Ising values $\nu=1$ and $\beta=1/8$, respectively, providing strong evidence that these transitions are of Ising type.

\subsection{Crossover}
\begin{figure}
    \includegraphics[width=1\linewidth]{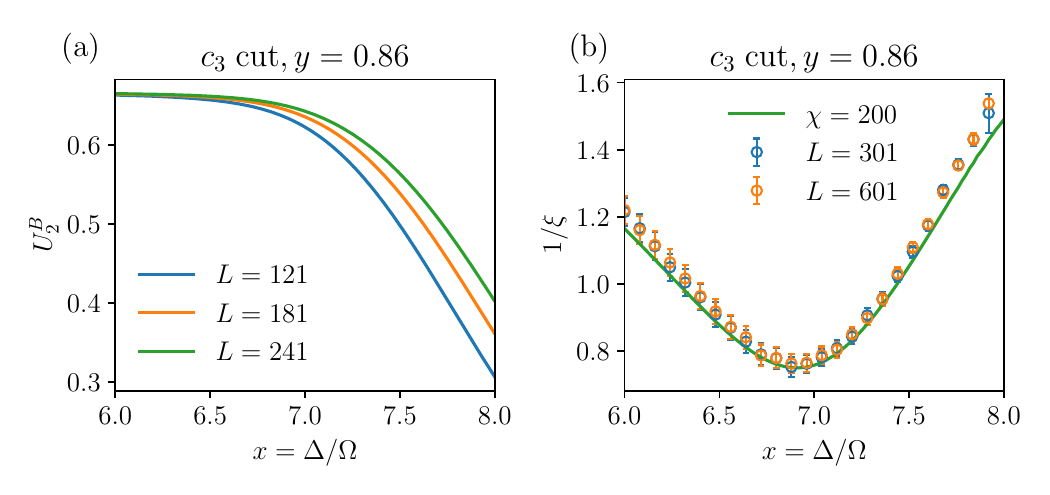}
    \caption{
    (a) Binder cumulant $U_2^B$ for different system sizes at $c_3$ cut ($y=R_b/d=0.86$).  
    (b) Inverse correlation length $1/\xi$ extracted from finite systems of size $L=301$ and $L=601$ by fitting the correlation functions $C_s(x)$. Error bars indicate one standard deviation. The green line shows the result obtained from VUMPS \cite{Zauner2018} with bond dimension $\chi=200$.
    }
    \label{crossover}
\end{figure}

The transition between $\mathbb{Z}_2 \otimes \mathbb{I}$ and $\mathbb{Z}_2 \otimes \mathbb{Z}_2$ is particularly interesting because it does not correspond to a conventional phase transition. As $R_d/d$ decreases, or equivalently as $\Omega$ increases, the B atoms, whose interactions are weaker, become disordered first. Upon further decreasing $R_d/d$, even the A atoms can no longer withstand the transverse field and also become disordered. The $\mathbb{Z}_2 \otimes \mathbb{I}$ phase therefore serves as an intermediate regime between the fully ordered and fully disordered phases. However, we note that, in contrast to the one-dimensional two-species Rydberg atom array, where two successive Ising phase transitions separate the fully ordered and fully disordered phases \cite{Lei2024}, in the present case the system first undergoes a crossover from the fully ordered phase to the $\mathbb{Z}_2 \otimes \mathbb{I}$ ordered phase, and only subsequently reaches the disordered phase via a single Ising phase transition. 

As shown in Fig.~\ref{crossover} (a), upon going from the $\mathbb{Z}_2 \otimes \mathbb{Z}_2$ ordered phase to the $\mathbb{Z}_2 \otimes \mathbb{I}$ ordered phase, the Binder cumulant for the B atoms does not exhibit any crossing among different system sizes. Furthermore, the inverse correlation length $1/\xi$ extracted along this cut remains finite and does not approach zero. Taken together, these observations indicate that the transition between the two $\mathbb{Z}_2$-ordered phases is a crossover rather than a genuine phase transition.

\subsection{$\mathbb{Z}_3\otimes\mathbb{Z}_3$ order to disorder transition}

\begin{figure}
    \includegraphics[width=1\linewidth]{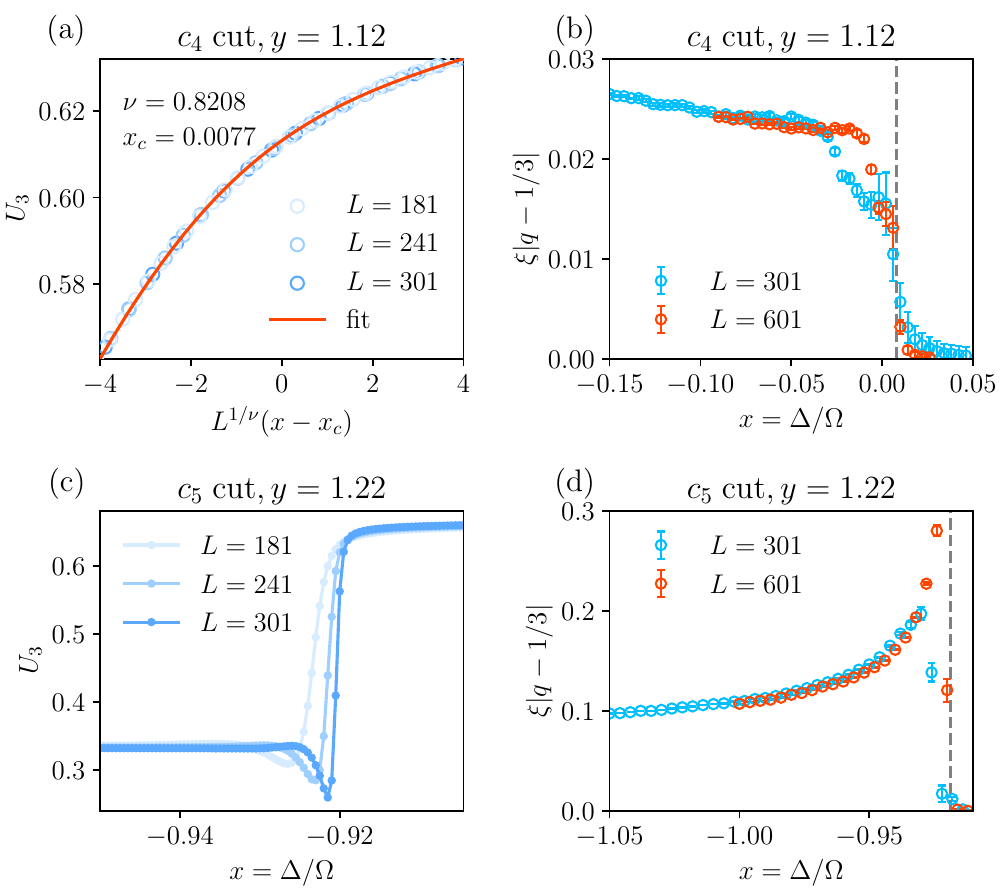}
    \caption{
    (a) Data collapse of the Binder cumulant along cut $c_4$, yielding a critical exponent $\nu \approx 0.82$. The data are fitted using a 20th-order polynomial. 
    (b) The phase transition indicator $\xi |q - 1/p|$ along cut $c_4$ approaches a finite, nonzero constant, indicating a chiral phase transition.
    (c) The Binder cumulant along cut $c_5$ exhibits a sudden drop, signaling the emergence of a floating phase.
    (d) The corresponding phase-transition indicator along cut $c_5$ diverges, indicating a KT transition. 
    }
    \label{Z3transition}
\end{figure}

\begin{figure}
    \includegraphics[width=1\linewidth]{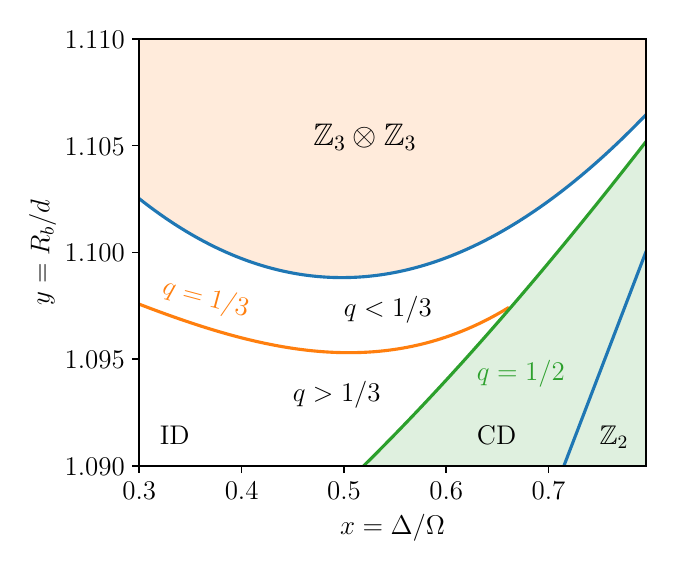}
    \caption{
    Illustration of the wave vector $q$ near the boundary between the disordered phase and the $\mathbb{Z}_3(\otimes \mathbb{Z}_3)$- and $\mathbb{Z}_2(\otimes \mathbb{Z}_2)$-ordered phases. The blue lines denote the phase boundaries extracted from the crossing points of the Binder cumulant for different system sizes. The green line marks the boundary between the commensurate disorder (CD) and incommensurate disorder (ID) phases. In the green (orange) shaded region, the wave vector takes the commensurate value $q = 1/2$ ($q = 1/3$). The orange line indicates the commensurate line within the ID region: above this line, the wave vector satisfies $q < 1/3$, while below it $q > 1/3$.
    }
    \label{wavevector3}
\end{figure}

The transition between the disordered phase and the $\mathbb{Z}_3 \otimes \mathbb{Z}_3$ 
ordered phase is a C-IC transition. Within the $\mathbb{Z}_3 \otimes \mathbb{Z}_3$ ordered phase, 
the ordering wave vector is commensurate and fixed at $q = 1/3$. 
On the disordered side, however, the wave vector varies continuously and is therefore 
generically incommensurate.
In principle, approaching the phase boundary along an incommensurate trajectory typically yields a chiral transition or a KT transition via an intervening floating phase. In contrast, approaching it along a commensurate trajectory is expected to produce a conformal critical point, which, if present here, should belong to the 3-state Potts universality class with critical exponent $\nu = 5/6 \approx 0.833$.

Unexpectedly, we do not find evidence for such a conformal critical point at the boundary of the $\mathbb{Z}_3 \otimes \mathbb{Z}_3$ ordered phase. As shown in Fig.~\ref{wavevector3}, the commensurate line $q = 1/3$ does not terminate at the ordered phase boundary. Instead, it ends at the boundary separating the CD region from the ID region. 
Consequently, only two scenarios remain for the transition between the disordered and $\mathbb{Z}_3 \otimes \mathbb{Z}_3$ ordered phases.

One scenario is a chiral phase transition, illustrated in Fig.~\ref{Z3transition} (a) and (b), where the extracted exponent $\nu$ deviates slightly from $5/6$ and the phase transition indicator approaches a finite, non-zero value near criticality. The other scenario involves crossing a floating phase, leading to a KT transition followed by a Pokrovsky-Talapov (PT) transition, as shown in Fig.~\ref{Z3transition} (c) and (d). 
In this case, the Binder cumulant exhibits a sharp drop, and the indicator $\xi |q - 1/3|$ diverges near the phase boundary, showing that the correlation length diverges faster than the wave vector mismatch. This behavior is characteristic of a KT transition, where the correlation length grows exponentially.

\begin{figure}
    \includegraphics[width=1\linewidth]{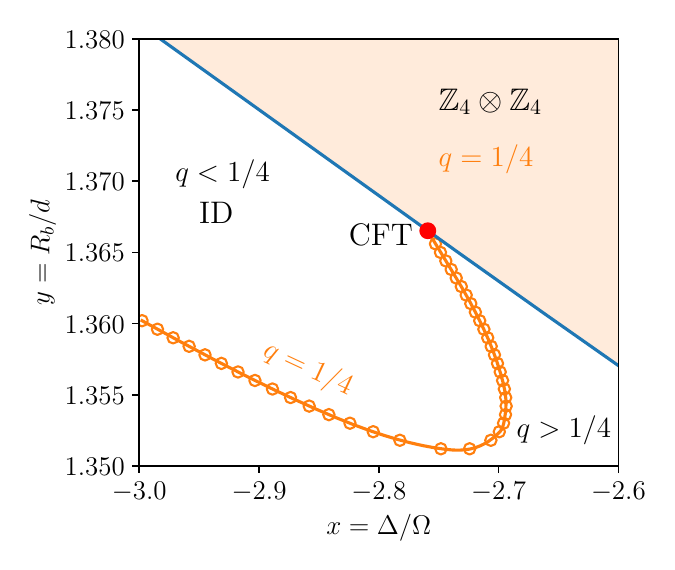}
    \caption{
    Illustration of the wave vector $q$ near the boundary between the disordered phase and the $\mathbb{Z}_4 \otimes\mathbb{Z}_4$ ordered phases.
    The blue lines indicate the phase boundaries extracted from the crossing points of the Binder cumulant for different system sizes.
    The orange shaded region corresponds to the commensurate regime where the wave vector is fixed at $q=1/4$.
    The orange line denotes the commensurate line within the incommensurate disordered (ID) region: above this line the wave vector satisfies $q<1/4$, while below it $q>1/4$.
    The CFT point, defined as the intersection of the phase boundary and the $q=1/4$ line, is highlighted by a large red dot.
    }
    \label{wavevector4}
\end{figure}

\begin{figure*}
    \includegraphics[width=1\linewidth]{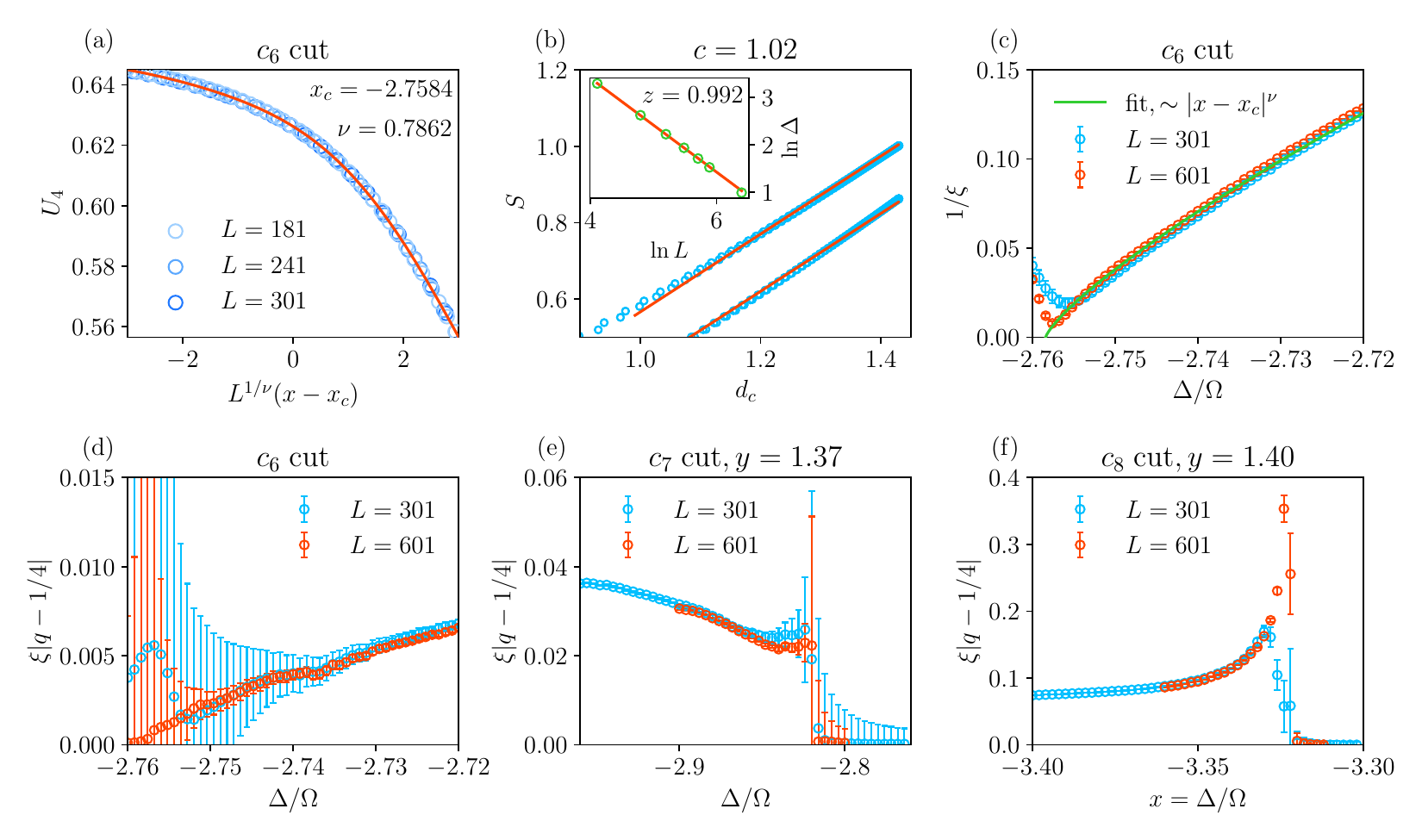}
    \caption{
    (a) Data collapse of the Binder cumulant along $c_6$ cut (CFT cut, $y=-0.1442x+0.9687$), yielding a critical exponent $\nu \approx 0.786$. The data are fitted using a 20th-order polynomial. 
    (b) Central charge $c$ and dynamical exponent $z$ (inset) are extracted at the critical point $x_c$. 
    (c) The correlation length at the disorder side show good agreement with the extracted exponent and $x_c$ from (a). 
    (d-f) Phase transition indicators along different cuts. 
    }
    \label{z4_transition}
\end{figure*}

\subsection{$\mathbb{Z}_4\otimes\mathbb{Z}_4$ order to disorder transition}

The transition from the $\mathbb{Z}_4 \otimes \mathbb{Z}_4$ ordered phase to the disordered phase is qualitatively similar to the $\mathbb{Z}_3 \otimes \mathbb{Z}_3$ case. The main difference is the presence of a CFT critical point located at the boundary of the $\mathbb{Z}_4 \otimes \mathbb{Z}_4$ ordered phase.

We find that the commensurate $q = 1/4$ line terminates at the boundary of the ordered phase, as shown in Fig.~\ref{wavevector4}. To further characterize the critical behavior, we perform a finite-size scaling analysis along the $c_6 cut$, which passes through the CFT point, as presented in Fig.~\ref{z4_transition} (a). From this analysis, we extract the critical exponent $\nu$ and the critical location $x_c$.
In Fig.~\ref{z4_transition} (b), we determine the central charge $c \approx 1$ and the dynamical critical exponent $z \approx 1$, confirming that this transition point is indeed described by a conformal field theory. The correlation lengths obtained from the correlation functions are consistent with the previously extracted $\nu$ and $x_c$, as shown in Fig.~\ref{z4_transition} (c).

Moreover, as shown in Fig.~\ref{z4_transition} (d), the indicator $\xi |q - 1/4|$ approaches zero near the critical point along the $c_6$ cut, signaling a conformal phase transition. This behavior is qualitatively different from that along the $c_7$ and $c_8$ cuts, shown in Fig.~\ref{z4_transition} (e) and (f). Along the $c_7$ cut, the indicator approaches a finite nonzero value, indicating a chiral transition. In contrast, along the $c_8$ cut, the indicator exhibits a clear divergence near the critical point, suggesting that the system crosses a floating phase. 


\subsection{Floating phases}
\begin{figure*}
    \includegraphics[width=1\linewidth]{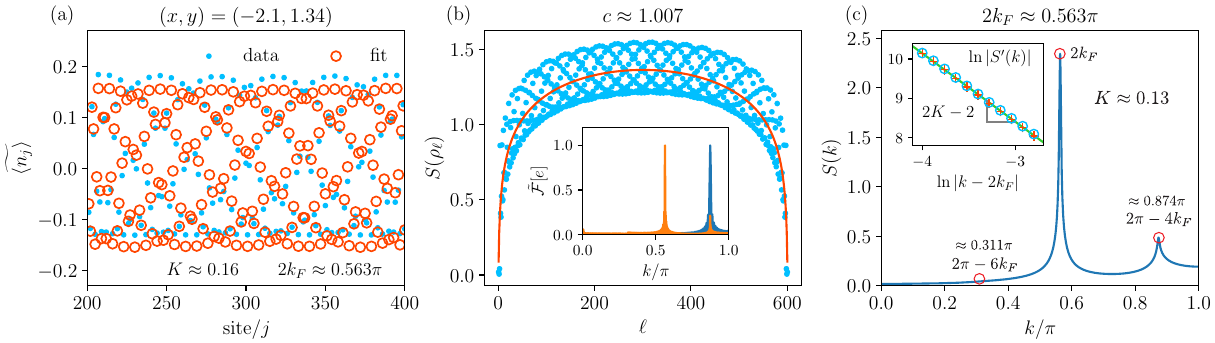}
    \caption{
    Data analysis of the floating phase at $(\Delta/\Omega, R_b/d)=(x,y)=(-2.1, 1.34)$, 
    obtained from a system of size $L=601$.
    (a) Friedel oscillations of the local excitation density $\tilde{n}_j$, fitted using Eq.\eqref{fridel_oscillation}. 
    Blue dots denote the DMRG data, while orange open circles indicate the fitting results.
    (b) Entanglement entropy $S(\rho_{\ell})$ for bi-partitions $\ell|(L-\ell)$, used to extract the central charge $c$. Here $\rho_{\ell}$ is the reduced density matrix of the first $\ell$ rungs. Blue dots and the orange solid line represent the DMRG data and the corresponding fit, respectively.
    Inset: Normalized Fourier transforms of the fitting residuals, denoted by $\widetilde{\mathcal{F}}[e]$. The blue and orange curves correspond to the residuals from fitting the excitation density, $e_n(j)=\tilde{n}_{j,\mathrm{data}}-\tilde{n}_{j,\mathrm{fit}}$, and the entropy oscillations, $e_S(\ell)=S(\rho_{\ell})-S_{\mathrm{fit}}(\rho_{\ell})$, respectively.
    (c) Structure factor $S(k)$, exhibiting two prominent peaks at $2k_F$ and $2\pi-2k_F$, as well as a much weaker peak at $2\pi-6k_F$.
    Inset: Determination of the Luttinger parameter $K$ using Eq.~\eqref{kspace_corr}. 
    The slope of the fitted line yields the exponent $2K-2$. Blue circles and orange crosses correspond to data points on the left and right sides of the $2k_F$ peak, respectively.
    }
    \label{floating}
\end{figure*}
Floating phases in Rydberg atom arrays have long been studied in the context of C-IC transitions \cite{Samajdar2018,Soto-Garcia2025}, and recently has been observed in experiments \cite{Zhang2025}.
In modern terms, floating phases are described as Luttinger liquids \cite{Giamarchi2003}, whose characteristics, such as the Luttinger parameter $K$ and the Fermi momentum $k_F$, depend on microscopic details, here controlled by $\Omega$ and $\Delta$. In contrast, universal features, including the central charge $c=1$ and power-law correlations, remain robust \cite{Michael2019}. 

The Luttinger liquids theory predicts the correlation function $C_s(x)$ in floating phase decays as 
\begin{equation}
    C_s(x) = \frac{K}{2\pi^2 x^2}+C\frac{\cos(2k_F x)}{x^{2K}}+\cdots, 
\end{equation}
where $C$ is a nonuniversal constant \cite{Giamarchi2003,Karrasch2012}. 
In momentum space, this translates to \cite{Karrasch2012}
\begin{equation}
    \frac{\mathrm{d} S(k)}{\mathrm{d} k} = \begin{cases}
        K/(2\pi) & k\to0,\\
        \sim (2k_F-k)^{2K-2} & k\to 2k_F. 
    \end{cases}\label{kspace_corr}
\end{equation}
where the small-$k$ limit reflects the long-distance behavior of the real-space correlations. 
Since this regime cannot be reliably accessed in finite systems, it is more appropriate to extract the Luttinger parameter $K$ from the second line of Eq.~\eqref{kspace_corr}. 

In the floating phase, open boundaries act as effective impurities and induce Friedel oscillations in the bulk. These oscillations are predicted to take the form \cite{Cea2024,White2002}
\begin{equation}
    \tilde{n}_j\propto \frac{\cos(qj+\alpha)}{[(L/\pi)\sin(\pi j/L)]^K}, \label{fridel_oscillation}
\end{equation}
where $q$ is the incommensurate wave vector and $\alpha$ is a phase shift. 
In principle, fitting the Friedel oscillations using Eq.~\eqref{fridel_oscillation} allows for an accurate determination of both the Luttinger parameter $K$ and the wave vector $q$. However, in practice, such fits can become unreliable when higher-harmonic oscillations are strong. In these cases, the structure factor provides a more robust approach to extracting $K$.

As an illustrative example, we analyze a representative point deep inside the floating phase at $(x,y)=(-2.1,1.34)$, shown in Fig.~\ref{floating}. As seen in Fig.~\ref{floating} (a), the fit to the density fluctuations $\tilde{n}_j$ is not ideal, with many data points deviating systematically upward from the fitted curve. We attribute this discrepancy to the presence of strong higher-harmonic oscillations. This interpretation is supported by the Fourier transform of the fitting residuals, which exhibits a single pronounced peak at $2\pi-4k_F$, as shown by the blue curve in the inset of Fig.~\ref{floating} (b). This indicates that higher-order oscillations play an essential role in this floating phase. Nevertheless, the fitting procedure still yields reasonable estimates of $K$ and $2k_F$, compared to Fig.~\ref{floating} (c).
A similar phenomenon is observed in the entanglement entropy oscillations. By fitting the uniform part of the entanglement entropy, we extract a central charge $c$ close to 1, indicating that the low-energy physics is governed by a single bosonic mode. Subtracting the uniform contribution, shown by the orange line in Fig.~\ref{floating} (b), reveals residual oscillations that closely resemble those of the density fluctuations. The Fourier transform of these residuals exhibits peaks at $2k_F$ and $2\pi-4k_F$, as shown by the orange curve in the inset of Fig.~\ref{floating} (b).

Finally, Fig.~\ref{floating} (c) displays the structure factor $S(k)$ of the same point. The dominant peak corresponds to the wave vector $2k_F$, while the $4k_F$ contribution, folded back into the first Brillouin zone as $2\pi-4k_F$, is also clearly visible. The $6k_F$ peak is present but significantly weaker. Using Eq.~\eqref{kspace_corr}, we extract the Luttinger parameter $K \simeq 0.13$. This estimate, obtained from the structure factor, is expected to be more reliable than that obtained from fitting the real-space density oscillations.

\subsection{Multi-critical point}

Numerical results indicate the presence of a multi-critical point in the phase diagram, located at the boundary between the $\mathbb{Z}_2 \otimes \mathbb{Z}_2$ and $\mathbb{Z}_3 \otimes \mathbb{Z}_3$ ordered phases. This multi-critical point is surrounded by an incommensurate disordered region and two distinct ordered phases, and serves as the common endpoint of the Ising transition line, the chiral transition line, and the first-order transition line.

To capture the competing ordered phases and the multi-critical behavior observed in the phase diagram, we propose a minimal field action in terms of two order-parameter fields. 
The real scalar field $\phi(x)\in\mathbb{R}$ represents an Ising-like $\mathbb{Z}_2$ order, while the complex field $\psi(x)\in\mathbb{C}$ describes a period-3 density-wave order. 
These fields enter the long-wavelength expansion of the rung density as 
\begin{equation}
    n_r(x) = n_0+(-1)^x\phi(x) + \mathrm{Re}[\psi(x)e^{i2\pi x/3}]+\cdots, 
\end{equation}
and the minimal symmetry-allowed effective action is 
\begin{eqnarray}
    \mathcal{S}[\phi,\psi] &=& \int\mathrm{d}\tau\mathrm{d}x\ (\mathcal{L}_{\phi}+\mathcal{L}_{\psi}+\mathcal{L}_{\phi\psi}), \\
    \mathcal{L}_{\phi} &=& (\partial_{\tau}\phi)^2+v_{\phi}(\partial_x\phi)^2+r_{\phi}\phi^2 +u_{\phi}\phi^4, \\
    \mathcal{L}_{\psi} &=& |\partial_{\tau}\psi|^2+v_{\psi}|\partial_x\psi|^2+r_{\psi}|\psi|^2 +u_{\psi}|\psi|^4 \nonumber\\
    &{}& + \lambda(\psi^3+\psi^{*3}) + i\alpha(\psi^*\partial_x\psi - \psi\partial_x\psi^*), \\
    \mathcal{L}_{\phi\psi} &=& g\phi^2|\psi|^2.
\end{eqnarray}
Here $\mathcal{L}_{\phi}$ describes a conventional Ising transition associated with the $\mathbb{Z}_2$ symmetry breaking, while $\mathcal{L}_{\psi}$ governs the behavior of the complex period-3 density-wave order parameter. The cubic term $\lambda(\psi^3+\psi^{*3})$ acts as a commensurate lock-in perturbation, selecting the three symmetry-related period-3 configurations and stabilizing the $\mathbb{Z}_3$ ordered phase.
The effective action also contains a symmetry-allowed linear derivative term $i\alpha(\psi^*\partial_x\psi-\psi\partial_x\psi^*)$. It breaks emergent Lorentz invariance and biases the propagation of phase fluctuations, driving the critical behavior toward a chiral universality class, as discussed in \cite{Fendley2004,Samajdar2018}.
The coupling term $g\phi^2|\psi|^2$ captures the competition between the $\mathbb{Z}_2$ and $\mathbb{Z}_3$ orders. For repulsive coupling $g>0$, coexistence of the two orders is energetically unfavorable, which generically leads to a first-order transition between the $\mathbb{Z}_2$ and $\mathbb{Z}_3$ ordered phases. The combined effects of Ising criticality, chiral criticality in the $\psi$ sector, and the competitive inter-order coupling is expected to give rise to a multi-critical point where the Ising transition line, the chiral transition line, and the first-order boundary intersect. We leave a detailed analysis of this effective action for future work.

\section{conclusion and outlook}
In this work, we have uncovered a rich phase diagram in a dual-species Rydberg atom ladder, where the interplay between competing interactions and ladder geometry gives rise to a variety of quantum phases and phase transitions beyond those realized in single-species systems. In particular, we identify extended floating phases with incommensurate correlations, a smooth crossover between distinct $\mathbb{Z}_2$-ordered regimes, and a multi-critical point at which Ising, chiral, and first-order transition lines intersect. These features highlight the qualitative impact of multi-component interactions, which cannot be captured by a single blockade constraint and lead to fundamentally new collective behavior.

Our results demonstrate that dual-species Rydberg systems provide a natural platform for realizing chiral criticality and multi-critical phenomena in one dimension. In contrast to single-species arrays, where the phase diagram is largely governed by a single interaction scale, the presence of multiple blockade radii introduces competing ordering tendencies and enables the emergence of crossover physics and nontrivial phase boundaries. The absence of a conventional three-state Potts critical point and its replacement by a chiral transition further emphasizes the distinct universality structure of the system.

From a theoretical perspective, it would be highly desirable to develop a more complete field-theoretical description of the multi-critical region. While the minimal coupled order-parameter theory proposed here captures the essential competition between $\mathbb{Z}_2$ and $\mathbb{Z}_3$ orders, a systematic analysis of its renormalization-group flow and the role of marginal operators remains an open problem. 
On the experimental side, the parameter regimes explored in this work are directly accessible in current Rydberg platforms with programmable tweezer arrays. The floating phases and crossover behavior can be probed through spatial correlation functions and structure-factor measurements, while the multi-critical point may be identified via finite-size scaling of observables and dynamical signatures. Extending these studies to higher-dimensional geometries, or to driven and dissipative settings, may reveal even richer non-equilibrium and topological phenomena.

Overall, our work establishes dual-species Rydberg atom arrays as a versatile setting for exploring competing orders, crossover physics, and multi-critical quantum behavior, opening new avenues for both theoretical and experimental investigations of strongly correlated quantum matter.

\begin{acknowledgments}
This work is supported by the National Key Research and Development 
of China (Grant Nos. 2021YFA1402001, 2021YFA0718304) and the National Natural Science Foundation of China (NSFC) (Grant Nos. 12375007, 12574295).
\end{acknowledgments}

\appendix
\section{Details of Correlation Function Fitting}

To extract the correlation length $\xi$ and the wave vector $q$ from finite-size DMRG data, we employ a two-step fitting procedure.

In the first step, we obtain a rough estimate of the correlation length, denoted by $\xi_0$, by fitting
\begin{equation}
    \ln \left|C(x)\sqrt{x}\right| \simeq -\frac{x}{\xi_0} + \ln |A_0| + C,
    \label{fitting_eq1}
\end{equation}
where the oscillatory factor $\cos(2\pi q x + \phi_0)$ in the correlation function $C(x)$ is neglected. Its effect is approximately absorbed into the constant term $C$.

In the second step, we refine this estimate by writing the correlation length as
\begin{equation}
    \xi = \xi_0 + \delta \xi,
\end{equation}
where $\delta \xi$ is expected to be small. We then consider the transformed quantity
\begin{equation}
    C(x)\sqrt{x}\, e^{x/\xi_0} \simeq A_0 \exp\!\left[x\left(\frac{1}{\xi_0} - \frac{1}{\xi_0 + \delta \xi}\right)\right] \cos(2\pi q x + \phi_0),
\end{equation}
and determine the parameters $A_0$, $\delta \xi$, $q$, and $\phi_0$ by minimizing the cost function
\begin{widetext}
\begin{equation}
    F(A_0, \delta \xi, q, \phi_0)
    = \sum_{x=a}^{b}
    \left|
    C(x)\sqrt{x}\, e^{x/\xi_0}
    - A_0 \exp\!\left[x\left(\frac{1}{\xi_0} - \frac{1}{\xi_0 + \delta \xi}\right)\right]
    \cos(2\pi q x + \phi_0)
    \right|^2,
\end{equation}
\end{widetext}
where $[a,b]$ defines the fitting window.

This two-step procedure yields accurate estimates of both the correlation length $\xi$ and the wave vector $q$.

\nocite{*}
\bibliography{ref}
\end{document}